\documentclass[aps,prl,showpacs,twocolumn]{revtex4}
\usepackage{amsmath,graphicx}
\begin{document}
\let\up=\uparrow
\let\down=\downarrow
\newcommand\ket[1]{|{\textstyle#1}\rangle}
\newcommand\bra[1]{\langle{\textstyle#1}|}
\newcommand\braket[1]{\langle{\textstyle#1}\rangle}
\newcommand\tilA{\widetilde{A}}
\newcommand\tilB{\widetilde{B}}
\newcommand\tilC{\widetilde{C}}
\newcommand\tilD{\widetilde{D}}
\newcommand\varD{\mathcal{D}}

\title{Andreev Bound States in the Kondo Quantum Dots Coupled to 
  Superconducting Leads}
\author{Jong Soo Lim and Mahn-Soo Choi}
\email{choims@korea.ac.kr}
\affiliation{Department of Physics, Korea University, Seoul 136-701,
  Korea}
\begin{abstract}
We have studied the Kondo quantum dot coupled to two superconducting
leads and investigated the subgap Andreev states using the NRG method.
Contrary to the recent NCA results [Clerk and Ambegaokar, Phys. Rev. B
61, 9109 (2000); Sellier et al., Phys. Rev. B 72, 174502 (2005)], we
observe Andreev states both below and above the Fermi level.
\end{abstract}
\maketitle

When a localized spin (in an impurity or a quantum dot) coupled to
BCS-type $s$-wave superconductors(S), two strong correlation effects
compete.  The superconductivity tends to keep the conduction electrons
in singlet pairs\cite{Tinkham96a}, leaving the local spin unscreened.
The spin state of the total wave function is thus a doublet.  The Kondo
effect tends to screen the local spin with the spins of the
quasiparticle excitations in the superconductor.  The total spin state
is thus a singlet\cite{Hewson93a}.  This competition gives rise to a
quantum phase transition from the doublet to singlet state, and the
transport properties are dramatically changed across this
transition\cite{Shiba69a,Glazman89a}.  For example, the Josephson
current through the quantum dot (QD) coupled to two superconducting
leads has a $\pi$-shift in its current-phase relation (a so-called
$\pi$-junction behavior) for the doublet state, while it has a
$0$-junction
behavior for the singlet state\cite{Rozhkov99a,Rozhkov00a,Clerk00b,ChoiMS04c,Siano04a,Siano04a-Erratum,ChoiMS05d,Siano05a,Sellier05a}.
In fact, all the previous
studies\cite{Rozhkov99a,Rozhkov00a,Clerk00b,ChoiMS04c,Siano04a,Siano04a-Erratum,ChoiMS05d,Siano05a,Sellier05a}
of the transition between the doublet and singlet state focused on the
current-phase relation $I_S(\phi)$ of the Josephson current.

However, there are another non-trivial issues, about the Andreev bound
states in such a system.  The issues are: (1) How many subgap Andreev
states are there?  (2) Are the subgap Andreev states true bound states
or quasi-bound states with finite level broadening?

Using the non-crossing approximation (NCA), \citet{Clerk00b}
investigated the close relation between the $0$-$\pi$ transition in
$I_S(\phi)$ and the Andreev states.  They
found that there is only one subgap Andreev state and that the Andreev
state is located below (above) the Fermi energy $E_F$ in the doublet
(singlet) state.  They provided an intuitively appealing interpretation
that in the doublet state, the impurity level well below $E_F$ is singly
occupied and due to the strong on-site interaction energy $U$ only
hole-like excitations are allowed. In the singlet state, a small
probability to find the impurity empty, electron-like excitations are
allowed.  This result was supported further by a more elaborated NCA
method by \citet{Sellier05a}.  As stressed by \citet{Clerk00b}, this
observation casts a strong contrast with the non-interacting case, where
bound states always occur in pairs (below and above
$E_F$)\cite{Beenakker92a,Beenakker92b}.  Moreover, the Andreev state has
a finite broadening.

To the contrary, the Hartree-Fock theory (HFT) \cite{Shiba73a} predicts two
Andreev states both below and above $E_F$ in the Kondo regime.  This was
in agreement with the numerical renormalization group (NRG) calculations
by \cite{Yoshioka00a}, who studied the Andreev states as a function of
the impurity level position.  Slave-boson mean-field theory (SBMFT)
\cite{Rosa} also predicts both
electron-like (above $E_F$) and hole-like (below $E_F$) Andreev
states.  Further, they both predict infinitely sharp Andreev states.
However, HFT and SBMFT are effectively non-interacting theory and may not
provide a strong argument against the NCA result. In a previous
work\cite{ChoiMS04c} we observed both Andreev states in the NRG result,
but the model had the particle-hole symmetry.

In this work, we report a systematic study of the issues
using the NRG method.  We find both electron-like and hole-like Andreev
states except in the deep inside the doublet phase (superconducting gap
even bigger than the hybridization).  We also provide a
supporting interpretation based on the variational wave functions
suggested by \citet{Rozhkov00a}.

The subgap Andreev bound states are important because
they are directly related to the transport properties of the
S/QD/S systems as in the recent
experiment\cite{Buitelaar02a,Buitelaar03a,vanDam06a,Jarillo-Herrero06a}.
Recent developments in
mesoscopic transport experiments may allow direct measurement of these
Andreev states through tunneling spectroscopy.  So far, the NCA/SNCA and
the NRG are the only methods which can treat rather systematically the
many-body correlations of the Anderson-type impurity coupled to
superconductors.  The discrepancy between the two methods
may motivate further theoretical efforts for better understanding of the
many-body states of the system.

\paragraph{Summary of the Results}
We consider a QD (or magnetic impurity) coupled to two superconducting
leads.  The Hamiltonian $H=H_C+H_D+H_T$ consists of three parts:
$H_C$ describes two, left ($L$) and right ($R$), BCS-like $s$-wave
superconductors with the superconducting gap $\Delta_{L(R)}$ and band
width $D$
\begin{equation}
H_C = \sum_{\ell=L,R}
\left[\sum_{k\sigma}\xi_{\ell k}
c_{\ell k\sigma}^\dag c_{\ell k\sigma} -
\left(\Delta_\ell c_{\ell k\up}^\dag c_{\ell k\down}+h.c.\right)
\right]
\end{equation}
We assume identical superconductors,
\begin{math}
\xi_{Lk} = \xi_{Rk} = \xi_k
\end{math}
and
\begin{math}
\Delta_L = \Delta_R^* = \Delta e^{+i\phi/2}
\end{math},
where $\phi$ is the phase difference between the two superconductors.
$H_D$ describes an Anderson-type single level in the QD
\begin{equation}
H_D = \sum_\sigma \epsilon_dd_\sigma^\dag d_\sigma
+ Un_\up n_\down
\end{equation}
with $n_\sigma=d_\sigma^\dag d_\sigma$.  $\epsilon_d$ is the
single-particle energy of the level and $U$ is the on-site interaction.
Here we consider particle-hole \emph{asymmetric} case
($\epsilon_d\neq-U/2$).  To compare our results directly with the NCA
results, we will mostly take $U=\infty$ by preventing double occupancy.
Finally $H_T$ is responsible for the the tunneling of electrons between
the superconductors and the QD
\begin{equation}
H_T = \sum_{\ell k\sigma}\left(
  V_{\ell k} c_{\ell k\sigma}^\dag d_\sigma + h.c.\right)
\end{equation}
For simplicity we will assume the symmetric junctions with tunneling
elements insensitive to the energy, $V_{Lk}=V_{Rk}=V$.  The broadening
of the level is given by
\begin{math}
\Gamma
= \pi\rho_L(E_F)|V_L|^2 + \pi\rho_R(E_F)|V_R|^2
= 2\pi\rho(E_F)|V|^2 \,.
\end{math}
For the calculation, we followed the standard NRG
method\cite{Wilson75a,Krishna-murthy80a,Krishna-murthy80b,Costi94a,Bulla98a}
extended to superconducting leads\cite{Yoshioka00a,Sakai89a,Satori92a}.

\begin{figure}
\centering
\includegraphics*[width=6cm]{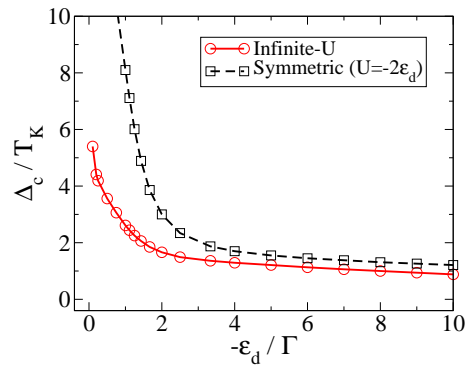}
\caption{(color on-line) Phase diagram of the Anderson impurity
  model with superconducting leads (for $\phi=0$).  The phase boundaries
  for the infinite-$U$ model (red solid line with circles) and for
  the particle-hole symmetric model (black dashed line with squares),
  respectively, have been calculated by the NRG method.}
\label{fig:3}
\end{figure}

There are two competing energy scales in the system.  The
superconductivity is naturally governed by the gap $\Delta$.  The Kondo
effect is characterized by the Kondo temperature $T_K$,
given by\cite{Haldane78a,Haldane78b,ChoiMS04c,ChoiMS05d}
\begin{equation}
T_K = \sqrt\frac{\Gamma W_0}{2}\,
\exp\left[\frac{\pi\epsilon_d}{2\Gamma}\,
  \left(1+\frac{\epsilon_d}{U}\right)
\right]
\end{equation}
where $W_0\equiv\min\{D,U\}$.  For $T_K\gg\Delta$, the ground state is
expected to be a singlet and the Josephson current is governed by the
Kondo physics.  In the opposite limit $T_K\ll\Delta$, the ground state
is a doublet and the transport can be understood perturbatively in the
spirit of the Coulomb blockade (CB) effect\cite{Glazman89a,Spivak91a}.
The transition happens at $T_K\sim\Delta$. See Fig.~\ref{fig:3}.

\begin{figure}
\centering
\includegraphics*[width=4cm]{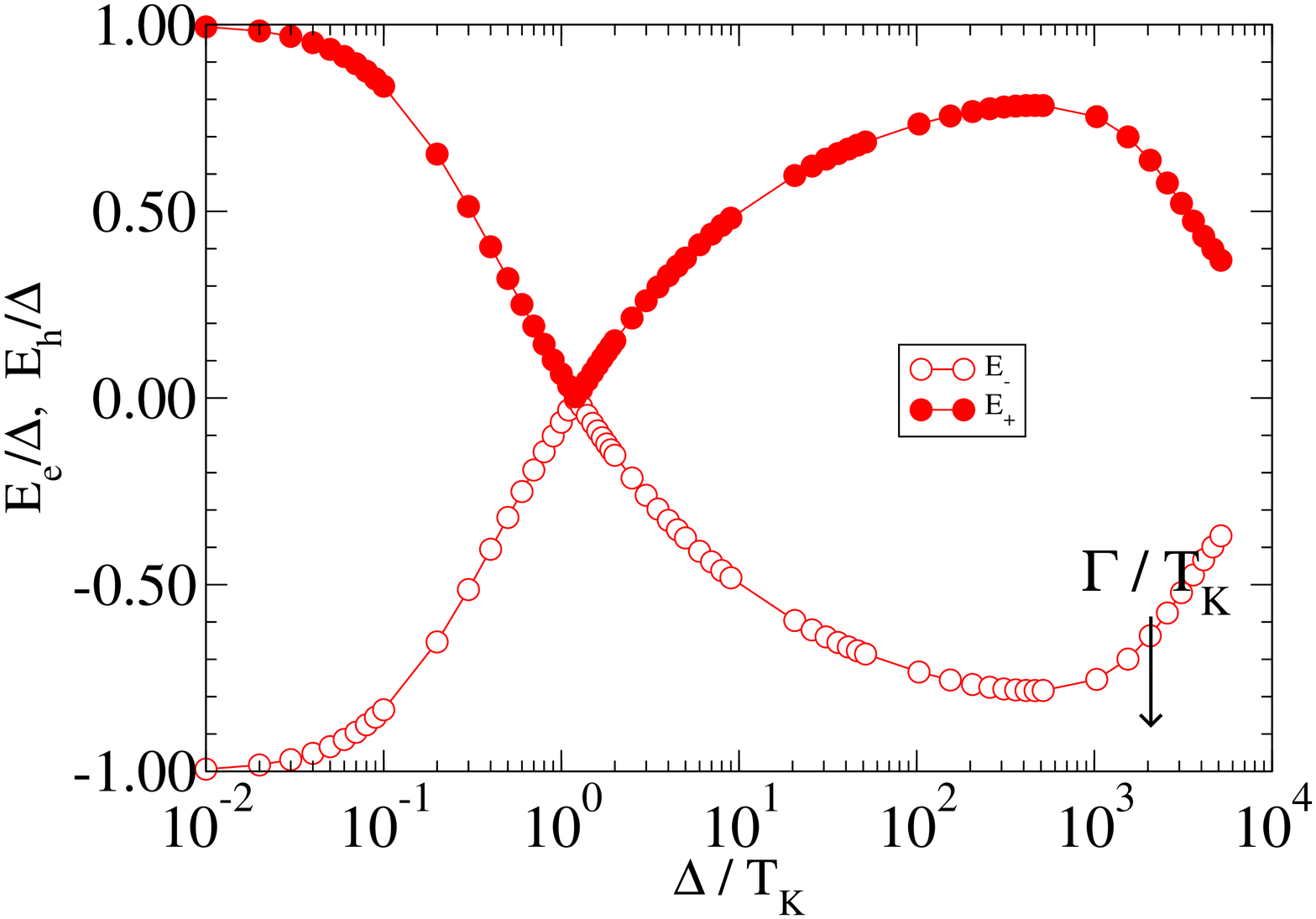}
\includegraphics*[width=4cm]{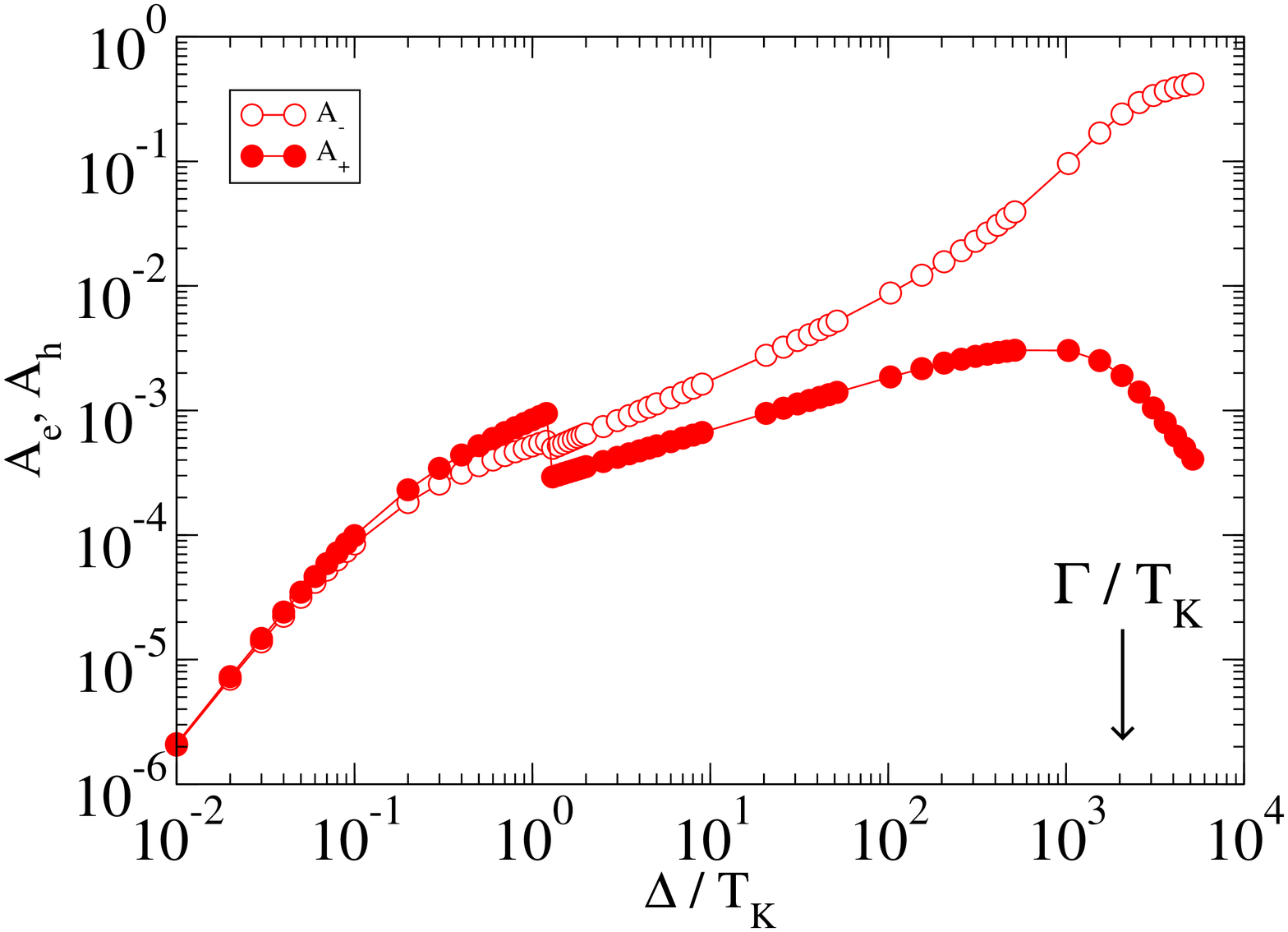} \\
\includegraphics*[width=4cm]{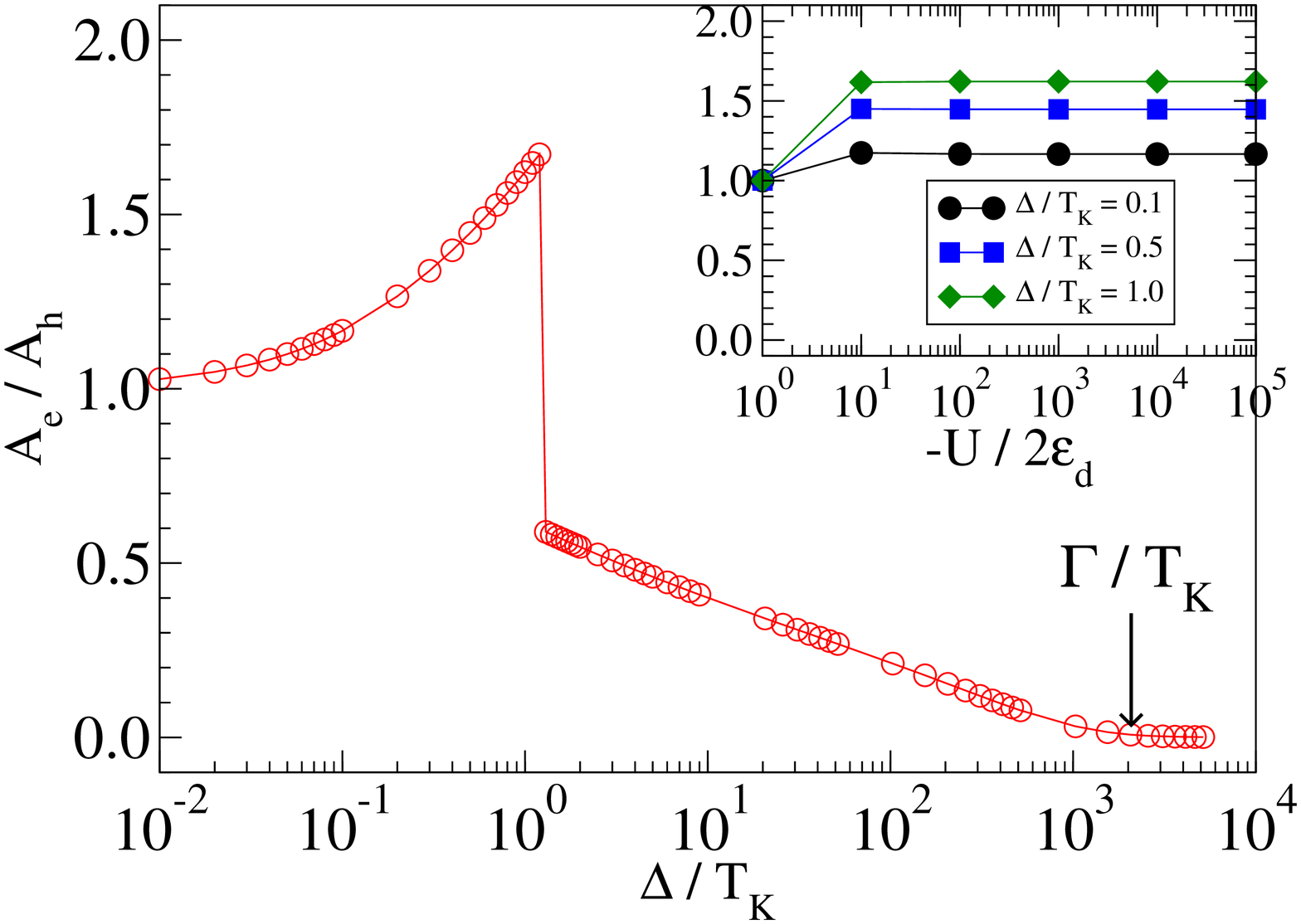}
\includegraphics*[width=4cm]{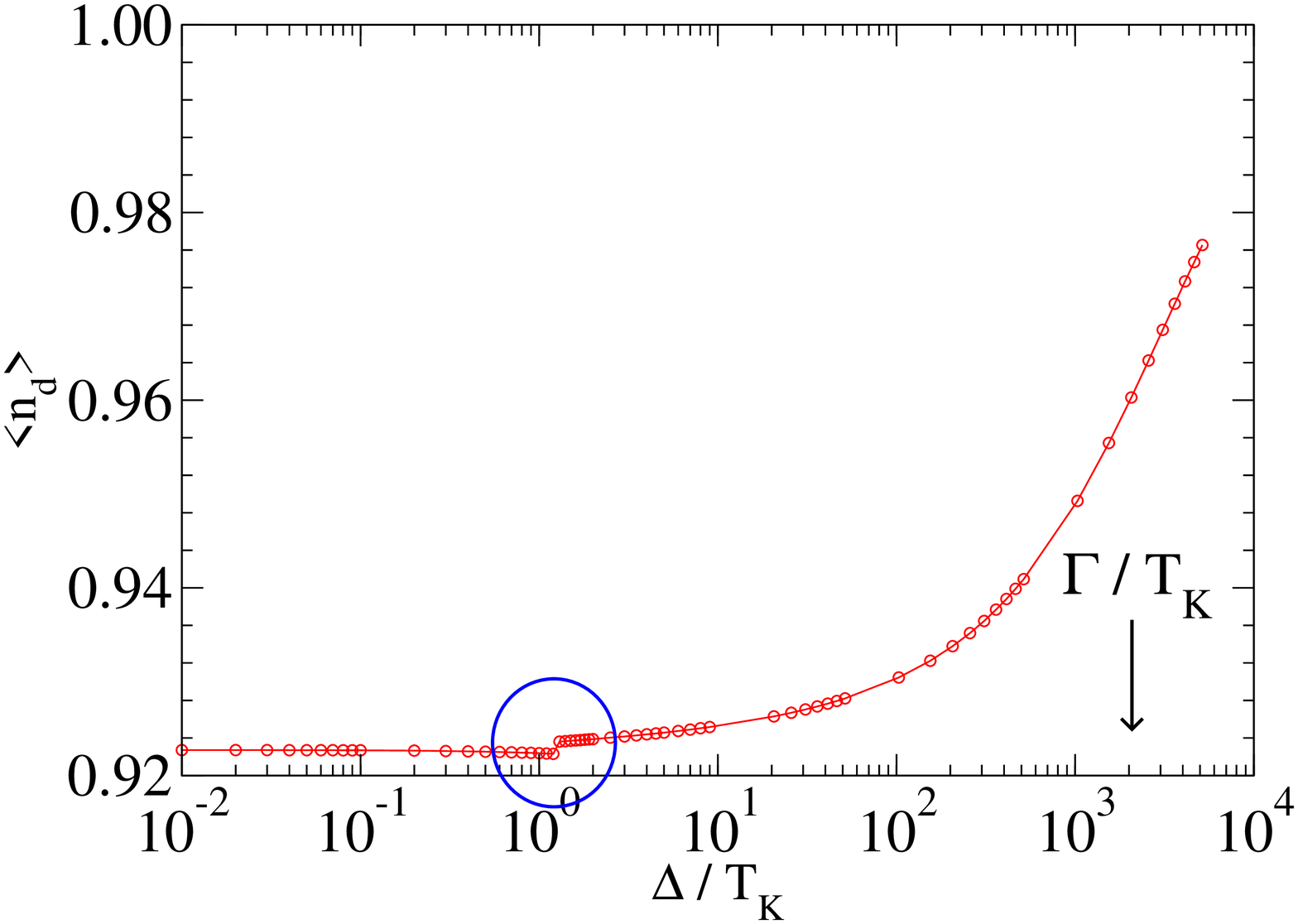}
\caption{(color on-line) The energies (a) and the corresponding spectral
  weights (b) of the subgap Andreev states for $U=\infty$ Anderson
  model.  (c) The ratio $A_e/A_h$ of the spectral weights of the Andreev
  states.  Inset: $A_e/A_h$ as a function of $-U/2\epsilon_d$.  (d)
  Average occupation of the QD level.  $\epsilon_d=-0.1D$, $U=\infty$,
  $\Gamma=0.02D$ ($T_K\approx 3.88\times 10^{-5}$).}
\label{fig:1}
\end{figure}

Figure~\ref{fig:1} summarizes the results.  Figure~\ref{fig:1} (a) shows
the positions, $E_e$ and $E_h$, of the subgap Andreev states for
$U=\infty$ and $\phi=0$.  We observe two Andreev states, below and above
$E_F$ are observed in a wide range of $\Delta/T_K$ (in particular on
both sides of the transition point $\Delta_c/T_K\sim 1$).
More important are Fig.~\ref{fig:1}
(b) and (c), the spectral weights $A_e$ ($A_h$) of the electron-like
(hole-like) Andreev states, defined by
\begin{equation}
G_{dd}^R(E) \approx \frac{A_p}{E - E_p + i0^+}
\end{equation}
near $E\simeq E_p$ ($p=e,h$).  Except for very large $\Delta$
($\Delta\gg\Gamma$), both $E_e$ and $E_h$ have order of magnitude the
same spectral weights.  These observations are consistent with the
occupation of the dot level shown in Fig.~\ref{fig:1} (d).  Unlike the
intuitive interpretation by \cite{Clerk00b}, the occupation does not
change much across the transition point, although there is a small jump
[emphasized in the blue circle in Fig.~\ref{fig:1} (d)].  The results in
Fig.~\ref{fig:1} remain qualitatively the same for finite $U$; an
example is show in the inset of Fig.~\ref{fig:1} (c).  Finite phase
difference (not shown in the figure) does not make any qualitative
change (about the existance of the Andereev states both above and below
$E_F$), either.

\paragraph{Variational Theory}
The main features of the results summarized above can be understood
\emph{qualitatively} in
terms of the variational wave functions\cite{Rozhkov00a}.
For the singlet state in the $U=\infty$ limit we take the trial function
of the form
\begin{multline}
\label{AndreevStates::eq:1}
\ket{S} = \Bigg\{A
  + \frac{1}{\sqrt{2}}\sum_{q\in L,R}B_q
  (\gamma_{q\up}^\dag d_\down^\dag - \gamma_{q\down}^\dag d_\up^\dag)
  \\\mbox{}
  + \sum_{qq'}C_{qq'}\gamma_{q\up}^\dag \gamma_{q'\down}^\dag
\Biggr\}\ket{0}
\end{multline}
with $C_{qq'}=C_{q'q}$.  For the doublet state we take
\begin{multline}
\label{AndreevStates::eq:2}
\ket{D_\up} = \Biggl\{
\tilA d_\up^\dag + \sum_q\tilB_q\gamma_{q\up}^\dag
+ \sum_{qq'}\tilC_{qq'}\gamma_{q\up}^\dag\gamma_{q'\down}^\dag
d_\up^\dag \\\mbox{}
- \frac{1}{\sqrt{3}}\sum_{qq'}\tilD_{qq'}\gamma_{q\up}^\dag
\left(\gamma_{q'\up}^\dag d_\down - \gamma_{q'\down}^\dag d_\up\right)
\Biggr\}\ket{0}
\end{multline}
with $\tilC_{qq'}=\tilC_{q'q}$ and $\tilD_{qq'}=-D_{q'q}$,
and analogously $\ket{D_\down}$.  The coefficients in the trial wave
functions are determined by minimizing
\begin{math}
E = \braket{\Psi|H\Psi}/{\braket{\Psi|\Psi}}
\end{math}
for $\ket\Psi=\ket{S}$ and $\ket\Psi=\ket{D_\sigma}$.

\begin{figure}
\centering
\includegraphics*[width=6cm]{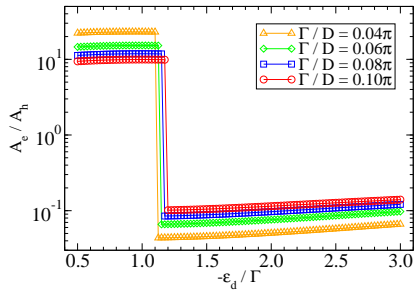}
\caption{(color on-line) Results from the variational calculations.
  Plotted is the Ratio $A_e/A_h$ of the spectral weights as a function
  of $-\epsilon_d/\Gamma$ for various values of $\Gamma$.
  $\Delta/D=0.1$.}
\label{fig:2}
\end{figure}

From the form of the trial wave functions, the spectral strength of the
hole-like Andreev state in the singlet phase depends on how much the QD
is occupied ($B_q$) in $\ket{S}$ and how much the QD is empty
($\tilB_q$) in $\ket{D_\sigma}$, namely, on the matrix element
\begin{equation}
\braket{D_\up|d_\down|S} = -\frac{1}{\sqrt{2}}\sum_q\tilB_q^*\tilB_q
\end{equation}
(up to normalization constant
\begin{math}
\sqrt{\braket{S|S}\braket{D_\up|D_\up}}
\end{math}).  The strength of the electron-like Andreev state in the
doublet phase also depends on $\braket{D_\up|d_\down|S}$.  Likewise, the
strength of the electron-like (hole-like) Andreev state in the singlet
(doublet) phase is determined by the matrix element
\begin{equation}
\braket{D_\up|d_\up^\dag|S} = \tilA^*A + \sum_{qq'}\tilC_{qq'}^*C_{qq'}.
\end{equation}
According to the NCA results\cite{Clerk00b,Sellier05a},
\begin{math}
\braket{D_\up|d_\down|S}
\end{math}
($\braket{D_\up|d_\up^\dag|S}$) should vanish in the singlet (doublet)
phase.  However, as shown in Fig.~\ref{fig:2}, neither of them
vanishes, and the spectral weights $A_e$ and $A_h$ are similar in order
of magnitude on both sides of the transition point, in agreement with
the NRG results.
We must point out that the agreement between the variational and NRG
results is only at a qualitative level.  The ratio $A_e/A_h$ from the
variational method is about 5 times bigger than the NRG result.
However, this is not surprising because the variational method is
limited in the regime $T_K$ not too large compared with $\Delta$; see
below.

There is another interesting point to be noticed in the variational wave
functions in Eqs.~(\ref{AndreevStates::eq:1}) and
(\ref{AndreevStates::eq:2}).  The lowest-energy solution to the
variational equation for $\ket{S}$ is well separated from the continuum.
This is also true for $\ket{D_\sigma}$.  This suggests that the subgap
Andreev state is a true bound state without broadening.  We will come
back to this discussion below.

\paragraph{Universality}
Since the singlet-doublet transition in the system is a true quantum
phase transition, the universality is also an important issue.  With
$\Delta$ and $T_K$ being the only two low energy scales in the system,
physical quantities should depend only on the ratio of $\Delta/T_K$ but
not on the details of the system.

\begin{figure}
\centering
\includegraphics*[width=8cm]{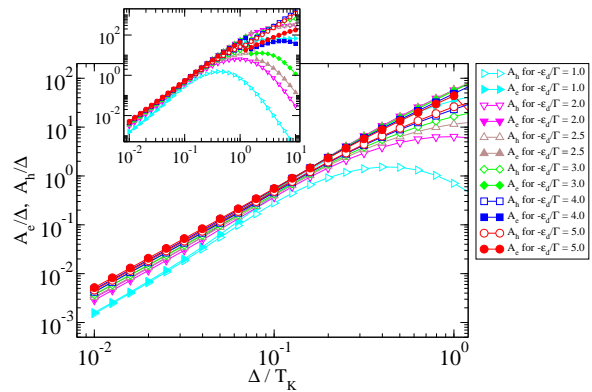}
\caption{(color on-line) Spectral weights $A_e$ ($A_h$) of the
  electron-like (hole-like) Andreev state as a function of $\Delta/T_K$
  at $\epsilon_d=\phi=0$.  Inset: the same in a wider range of
  $\Delta/T_K$ including the doublet phase.}
\label{fig:4}
\end{figure}

In Fig.~\ref{fig:4}, we plotted the normalized spectral weights
$A_{e(h)}/\Delta$ as a function of $\Delta/T_K$ for various values of
$\epsilon_d/\Gamma$.  We observe
that the curves of $A_{e(h)}$ overlap each other almost completely in
the Kondo regime ($\Delta\ll T_K$) except for cases close to the
mixed-valence regime ($|\epsilon_d|\lesssim\Gamma$).  The deviation from
the universal behavior in the mixed-valence regime
($-\epsilon_d\lesssim\Gamma$) is not surprising because of
strong charge fluctuations in the regime.
This is also indicated the phase diagram in Fig.~\ref{fig:3}: Close to
the mixed-valence regime ($-\epsilon_d/\Gamma\lesssim1$), $\Delta_c/T_K$
becomes larger.

More interestingly, the universal curve in Fig.~\ref{fig:4} can fit to a
simple effective non-interacting model.  On a non-interacting resonance
level coupled to two superconducting reservoirs, the spectral weights of
the Andreev states are given by
\begin{equation}
\frac{A_{e(h)}}{\Delta} = \frac{(1-\omega_0^2)}{\varD'(\omega_0)\varD(\omega_0)}
\left[z\left(1+\frac{\Gamma}{\sqrt{1-\omega_0^2}}\right)
  \pm \epsilon_d\right] \,,
\end{equation}
where
\begin{equation}
\varD(z) = z\left(\sqrt{1-z^2}+\Gamma\right)
- \sqrt{\epsilon_d^2(1-z^2) + \Gamma^2\cos^2(\phi/2)} \,.
\end{equation}
At the resonance ($\epsilon_d=0$) in the limit $\Gamma\gg\Delta$,
the expression is reduced to
\begin{equation}
\label{AndreevStates::eq:3}
\frac{A_{e(h)}}{\Delta} \approx 2\frac{\Delta^2}{\Gamma^2} \,.
\end{equation}
Kondo correlated state behaves like a Fermi liquid.  Naturally, if the
reservoirs are normal metal, the Kondo resonance can be regarded in
effect as a non-interacting resonance level at the Fermi energy $E_F$.
In other words, many physical properties described pretty well by the
effective impurity Green's function
\begin{equation}
G_d(z) = \frac{T_K/\Gamma}{z + iT_K}
\end{equation}
with $T_K$ playing the role of the level broadening.  In the previous
work\cite{ChoiMS04c} with particle-hole symmetry, it was demonstrated
that this may also work for superconducting reservoirs.  Indeed, the
spectral weights in Fig.~\ref{fig:4} fits very well to
\begin{equation}
\frac{A_{e(h)}}{\Delta} \sim \frac{\Delta^2}{T_K^2}
\end{equation}
to be compared with Eq.~(\ref{AndreevStates::eq:3}).

\begin{figure}
\centering
\includegraphics*[width=6cm]{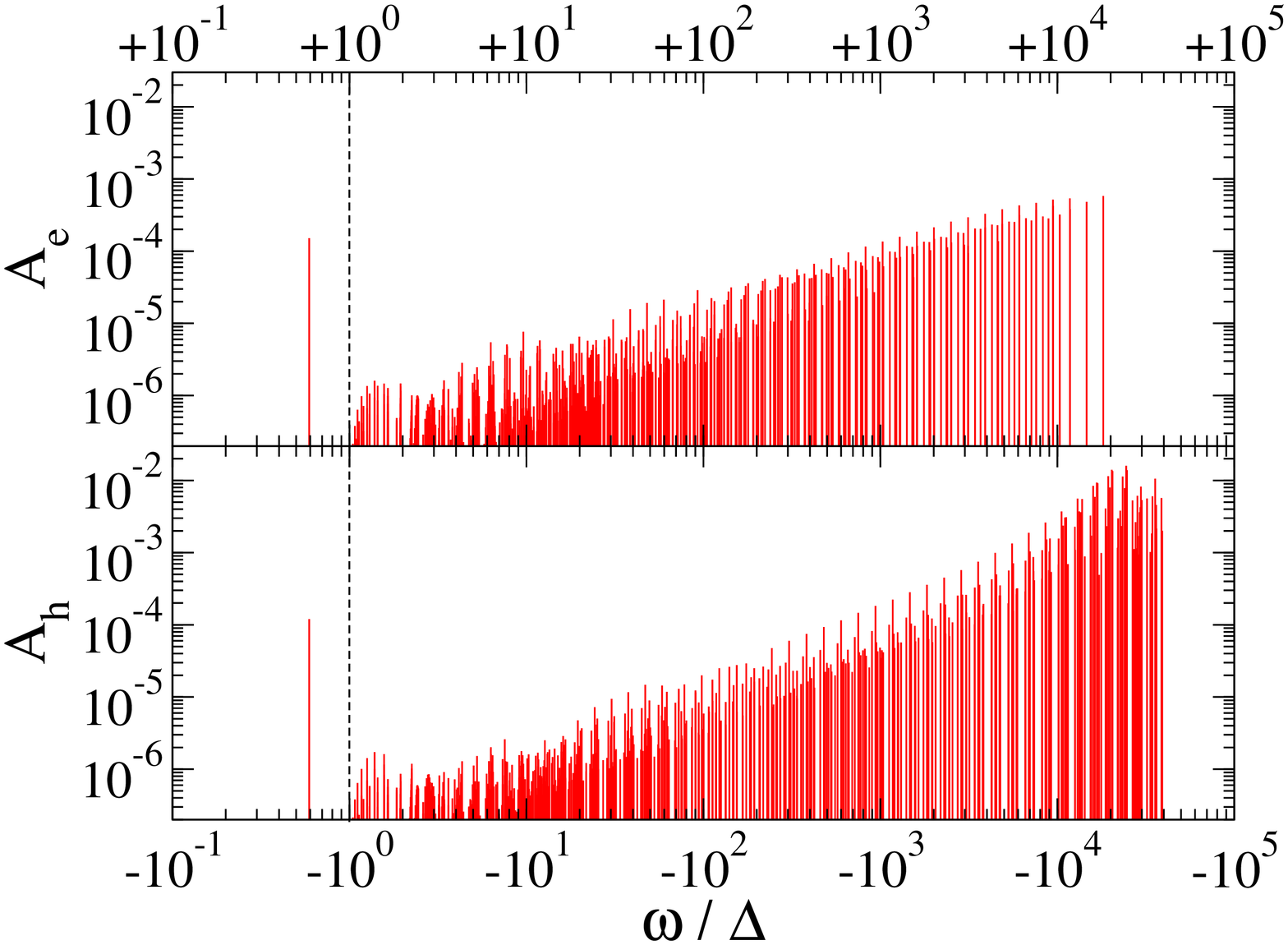}
\includegraphics*[width=6cm]{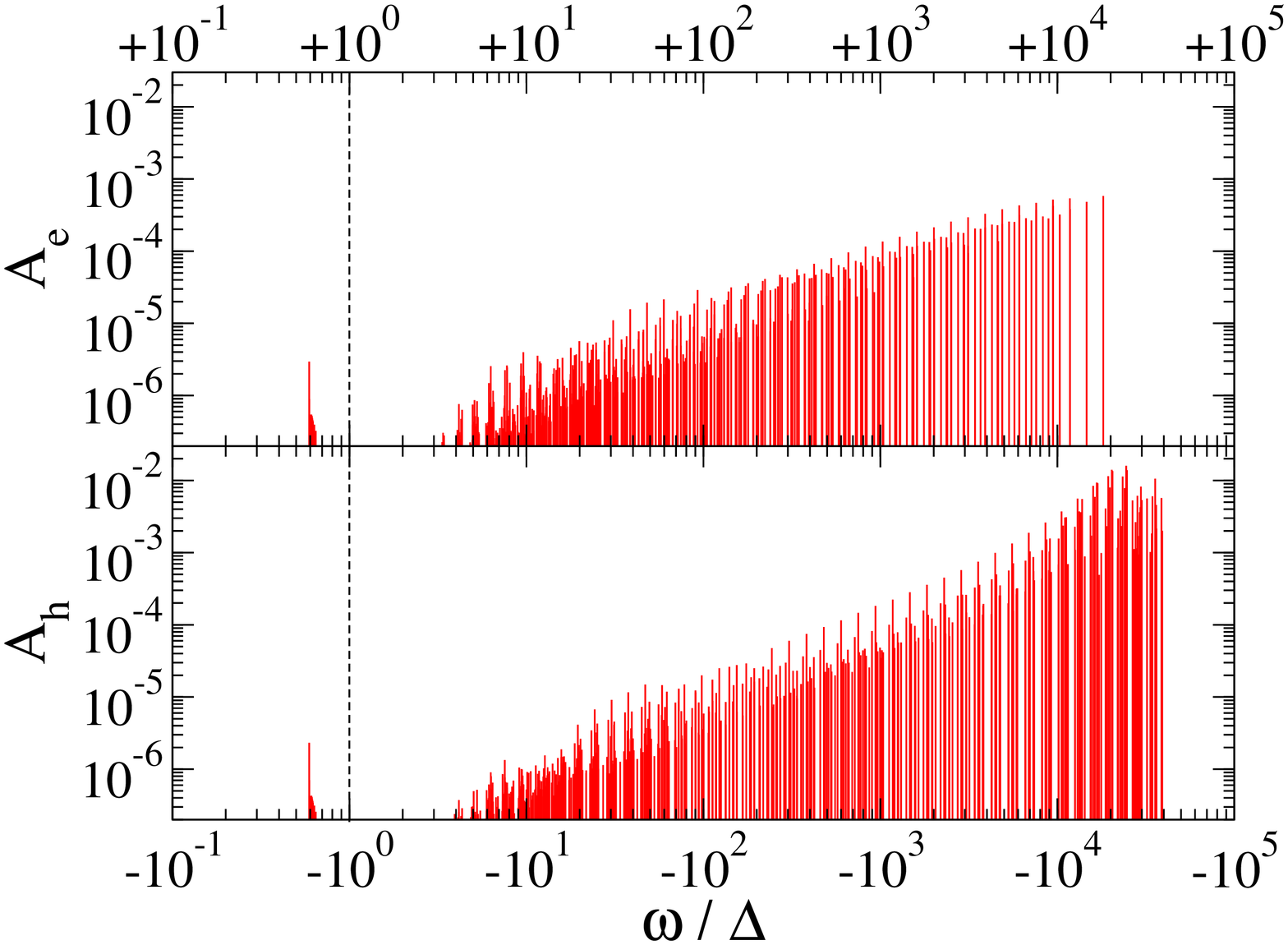}
\caption{(color on-line) Raw data of the spectral weights of the discrete
  energy levels from the NRG calculation (a) at $T=0.1\Delta$ and (b) at
  $T=0.5$.}
\label{fig:5}
\end{figure}

\paragraph{True bound state}
Finally, we address whether the subgap Andreev state is a true bound
state.  \citet{Clerk00a} and \citet{Sellier05a} found finite broadening of
the Andreev states.  This may come from the finite temperature effects.
The NCA cannot goes down to temperatures much lower than the
Kondo temperature, and they worked at rather high
temperatures\cite{Clerk00b,Sellier05a}.  The spectrum from the NRG
calculation is inherently discrete\cite{Bulla01a}, and it is not easy to
make a definite conclusion.  However, as shown in Fig.~\ref{fig:5}, the
subgap states are well separated from the continuum parts up to
temperatures as high as the energy of the subgap states. At temperatures
higher than the energy of the Andreev states, it is accompanied by other
small spikes.  It suggests that the subgap Andreev states are true bound
states.

\paragraph{Conclusion}
We have studied the Kondo quantum dot coupled to two superconducting
leads and investigated the subgap Andreev states using the NRG method.
Contrary to the recent NCA results\cite{Clerk00b,Sellier05a}, we observe
Andreev states both below and above the Fermi level.

%%%
Special thanks to A.~Clerk for the helpful discussions and for initially
drawing our attention to the issues. We also thank W. Belzig for comments.
This work was supported by the SRC/ERC program (R11-2000-071), the
KRF Grant (KRF-2005-070-C00055), the SK Fund, and the KIAS.

%%% Reference
\bibliography{AndreevStates}
% \bibliography{aliases,science,mathematics,physics,conmat,quaphy,staphy,%
%   opus,moreref}

\end{document}